\documentclass[manuscript,screen]{acmart}

\AtBeginDocument{%
  }

\setcopyright{none}
\renewcommand\footnotetextcopyrightpermission[1]{}

\usepackage{todonotes}
\usepackage{tcolorbox}
\usepackage{pgfplots}
\pgfplotsset{compat=1.18}

\begin{document}

\title[AI Companions for Young Adult Caregivers]{AI Companions for Young Adult Caregivers: Practical Advice, Emotional Depth, and the Limits of Metric Evaluation}

%%
%% Authors: one line per author, institution name only.
%% acmart requires a country for every affiliation, so \instonly passes an
%% invisible one (\relax); obeypunctuation stops acmart adding a comma before it.
\newcommand{\instonly}[1]{\affiliation[obeypunctuation=true]{\institution{#1}\country{\relax}}}

%% The first two authors contributed equally. Both get a star (*).
\author{Trizal Garg}
\instonly{BITS Pilani}

\author{Xuelong Fu}
\instonly{University of Cambridge}

\author{Dhruv Shah}
\instonly{BITS Pilani}

\author{Jin Huang}
\instonly{University of Cambridge}

\author{Shrigopal Maheshwari}
\instonly{BITS Pilani}

\author{Atiksh Aggarwal}
\instonly{BITS Pilani}

\author{Yash Sinha}
\instonly{BITS Pilani}

\author{Dhruv Kumar}
\instonly{BITS Pilani}

\author{Siddharth Mehrotra}
\instonly{BITS Pilani}

%% First-page footnote listing the authors and institutions.
\authorsaddresses{Authors' Contact Information: Trizal Garg, BITS Pilani; Xuelong Fu, University of Cambridge; Dhruv Shah, BITS Pilani; Jin Huang, University of Cambridge; Shrigopal Maheshwari, BITS Pilani; Atiksh Aggarwal, BITS Pilani; Yash Sinha, BITS Pilani; Dhruv Kumar, BITS Pilani; Siddharth Mehrotra, BITS Pilani.}

%%
%% Running head author list.
\renewcommand{\shortauthors}{Garg et al.}
\hypersetup{linkcolor=black}
\begin{abstract}
Young adult caregivers (YACs) are individuals aged 18 to 25 currently taking care of a family member, partner,
or friend facing chronic illness. AI companions designed specifically for YACs remain underexplored in prior works. We present Carebot, it uses an open-weight LLM (Gemma 3-27B) as its base model, on which we compared fine-tuning, retrieval-augmented generation (RAG), and a combination of both. The fine-tuning was performed based on an existing dataset of conversations with YACs \& a psychologist. Three well-being experts evaluated these configurations and preferred the fine-tuned model for the warmth of its responses and its actionable advice tailored to YACs. Based on their choice, we adopted the fine-tuned model as Carebot’s final configuration and used it for all subsequent evaluations. An LLM judge preferred this Carebot's responses over the base model's in 85\% of evaluated pairs. We then conducted a user study with 90 YACs, half of whom interacted with the base model and half with Carebot. The user study revealed a discrepancy between metric evaluation and actual user perception. Participants who used Carebot were more likely to characterize its replies as overly practical and brief, noting that it favored advice over emotional support. Conversely, participants who used the base model indicated that its responses were more emotionally oriented than practical. Although, fine-tuning made Carebot’s replies more practical, but they were not emotionally supportive for users. We argue that support tools for YACs require situation-sensitive emotional depth and aligned evaluation frameworks to study their effectiveness.
\end{abstract}

% Generate the CCSXML block at https://dl.acm.org/ccs/ccs.cfm and paste it here.


\maketitle

\section{Introduction}\label{sec:intro}
Family members provide much of the day-to-day care for relatives who live with illness, disability or frailty.
This care often affects the caregiver's own health, finances and time .
Researchers and services have responded with support for caregivers, such as information programs, respite care and peer support groups \cite{adelman2014caregiver}.
More recently, this support has moved to chatbots based on LLMs, which generate fluent replies to open questions \cite{cheng2025pdc30,parmanto2024reliable}.
These caregiving tools, however, target caregivers of all ages.
Young adult caregivers (YACs) are people aged 18 to 25 who provide regular care to a family member, partner,
or friend \cite{levine2005young,dang2024young}.
They take on this role while they finish their education, start work and build their own adult lives \cite{dang2024young}.
This group of people, often faces emotional, mental, and practical challenges that are not always
visible or well-supported.
In a survey of 295 young adult carers aged 14 to 25 in the United Kingdom, almost half reported a mental health problem \cite{becker2019young}.
Dang's work with student YACs in the Netherlands found that they perceived slightly less balance in life than students without caregiving duties \cite{dang2024perceived}.

A recent review of digital tools for young caregivers found only six studies and limited effects on mental health \cite{shankar2026effectiveness}.
We found no study that builds a generative AI companion for YACs.
This gap matters because a general-purpose chatbot can misjudge what a young caregiver needs.
One way to catch such mismatches is to evaluate a companion before it reaches users.
Common method is review by LLM-as-a-judge, in which a second LLM scores or compares replies \cite{zheng2023judging}.
Neither method involves YACs, so it is not known whether these checks predict how YACs experience a companion.

We present Carebot, an AI companion for YACs, using Gemma-3-27B.
It uses two techniques, fine-tuning (FT) and retrieval-augmented generation (RAG), to tailor the model to YACs.
We compared four configurations of the system, which were the base model, FT alone, RAG alone, and RAG combined with FT.
Human experts reviewed a sample of responses from each configuration and based on that review FT alone was selected.
Carebot, the final version tested with YACs, is the FT alone configuration.
The fine-tuning dataset consists of 10,000 synthetic conversations between a caregiver and a supportive companion, generated with GLM 5.2. To guide the generation, the model was given real caregiver transcripts as few-shot examples. The conversations balance gender and span 20 countries, 23 care-recipient conditions and the ten caregiving categories of Dang’s taxonomy \cite{dang2024young}.
We ran a user study with 90 YACs.
Half of them interacted with the base model and half of them with Carebot and then answered a questionnaire about their experience.
We also did an metric evaluation where we reviewed the replies manually, and alongside this we used an LLM judge that compared replies from both the base and Carebot.
We then compared this metric evaluation with the user-study.

In the metric evaluation, human experts preferred Carebot’s replies over those of the base model because, on 500 synthetic held-out examples, Carebot scored higher than base on three metrics out of five, namely BERTScore (0.576), BLEU (0.014) and ROUGE-L (0.195). An LLM judge (GLM 5.2) also favored Carebot’s replies in 85\% of decisive pairwise verdicts. The YACs who used Carebot were more likely to describe its replies as practical. However, the participants who used the base model more often described its replies as polite but not actionable. This work makes three contributions.

\begin{itemize}
  \item We present Carebot, an AI companion for YACs built on Gemma-3-27B (Section~\ref{sec:design}).
  \item We compare four configurations that combine FT and RAG. The comparison shows how FT and RAG changed replies as judged by human experts (Section~\ref{sec:design}).
  \item We report a between-subjects user study with YACs aged 17 to 24 ($n = 90$) to study the effectiveness of Carebot (Sections~\ref{sec:study} and~\ref{sec:results-study}).
\end{itemize}

\section{Related Work}\label{sec:related}

\subsection{Young Adult Caregivers and Their Support Needs}\label{sec:rw-yac}

Levine et al. gave one of the first accounts of YACs as a distinct group \cite{levine2005young}.
They reanalyzed two national caregiver surveys from the United States and focused on caregivers aged 18 to 25.
They estimated that YACs make up 12 to 18\% of adult caregivers.
Over half of these YACs were male, and their mean age was 21.
The YACs in these surveys also reported unmet needs, including needs for information.

Becker and Sempik surveyed 295 young adult carers aged 14 to 25 in the United Kingdom about their health, education and work \cite{becker2019young}.
Almost half of the respondents reported a mental health problem.
Lacey et al. used five rounds of a British household panel to follow young adults over time \cite{lacey2023inequalities}.
Those aged 16 to 29 who became caregivers reported fewer close friends in the short term.
Chevrier et al. reviewed 23 studies of young adult carers \cite{chevrier2022young}.
They found that studies identify YACs in different ways, which makes prevalence estimates vary widely.
They also found few or no support services designed for this age group.

Dang's doctoral thesis gives a broad account of YAC support needs \cite{dang2024young}.
The thesis includes a systematic review and a survey of life balance and well-being.
It also includes a study of the positive aspects of caregiving, and interviews and usability tests of a web-based support tool.
In the survey, student YACs in the Netherlands reported slightly less balance in life than students without caregiving duties \cite{dang2024perceived}.
In the interviews, 13 student YACs said they wanted professionals who listen without judging them \cite{dang2024exploring}.
They also wanted more information on how to give emotional support to the person they care for.
They were neutral about using the web-based tool in the future.
The thesis concludes that support should address both the difficulties and the rewards of caregiving \cite{dang2024young}.
Thomas studied young carers in New Delhi and Mumbai through interviews and a school survey \cite{thomas2025hidden}.
The study found that some children in India provide intensive care with little recognition or support.

This work differs from these studies in two ways.
First, most of the studies above come from the United States and Europe, and several cover wider age ranges, from 14 to 29.
Our user study included participants from across India.
Second, these studies describe YACs and their needs, but none of them has tested a conversational AI system for this group.

We draw on Dang's thesis even though it studied students in the Netherlands.
The thesis focuses on YACs specifically and examines their support needs from several angles.
Studies from other countries suggest that young carers face similar difficulties wherever they live.
Becker compared research and policy on young carers in the United Kingdom, Australia, the United States and Sub-Saharan Africa \cite{becker2007global}.
He concluded that young carers have much in common irrespective of where they live.
Lewis et al. surveyed 2{,}099 carers aged 15 to 17 in six European countries, including the Netherlands \cite{lewis2022first}.
In each country, carers reported lower well-being than their peers without caring duties.
Most carers also reported little formal support for their caring role.
Raji et al. reviewed 20 studies of young carers in low- and middle-income countries \cite{raji2025experiences}.
We therefore treat the needs that Dang identified as a reasonable starting point for YACs in India.
These needs include professionals who listen without judgment and information on how to give emotional support.

\subsection{Conversational Agents for Emotional and Caregiving Support}\label{sec:rw-agents}

Fitzpatrick et al. ran a randomized controlled trial of Woebot, a conversational agent that delivers cognitive behavioral therapy \cite{fitzpatrick2017delivering}.
Woebot follows scripted rules and does not use an LLM.
Later work studied companions built on LLMs.
Ma et al. analyzed posts and comments from an online community of Replika, an LLM-based companion app which does not cater to YAC's\cite{ma2023understanding}.
They reported problems with consistency, memory and harmful content, and some described overdependence.
Laestadius et al. analyzed 582 posts by Replika users and described how emotional dependence on the chatbot harmed some users' mental health \cite{laestadius2024too}.
Maeda and Quan-Haase argued that human-like features, such as personal pronouns and affirming language, lead users to attribute agency to chatbots and trust them \cite{maeda2024when}.
They described the resulting one-sided relationship as parasocial and linked it to harms such as the disclosure of sensitive data.
Sharma et al. took a different approach and used AI to assist human peer supporters \cite{sharma2023human}.
Their system gave supporters feedback on how empathic their replies were.
In a randomized trial with 300 supporters on a peer support platform, it raised conversational empathy by 19.6\% overall.

A small set of studies built LLM tools for caregivers.
Cheng and Ng grounded a GPT-4o chatbot in a guidebook on dementia caregiving \cite{cheng2025pdc30}.
Twenty-one family caregivers used it for two weeks and rated it easy to use and helpful.
Parmanto et al. fine-tuned small open models and connected them to a knowledge base on dementia caregiving \cite{parmanto2024reliable}.
Their caregiving language model outperformed GPT-3.5 on the metrics they reported.
Jo et al. studied CareCall, an LLM system that makes check-in calls to socially isolated people \cite{jo2023understanding}.
Their interviews with 34 users, operators and developers showed that the system eased loneliness.
It also led to tensions among stakeholders about what the system should do.
Shankar et al. reviewed digital tools for caregivers aged 25 or younger \cite{shankar2026effectiveness}.
They found only six studies, limited effects on mental health, and much lower adherence to self-guided tools than to structured programs.

Miner et al. tested four smartphone assistants and found that they responded inconsistently to crisis statements \cite{miner2016smartphone}.
None of the assistants referred users to a helpline for depression.
Moore et al. tested LLMs and commercial therapy bots against clinical standards of care \cite{moore2025expressing}.
The models expressed stigma toward some conditions and responded inappropriately to some crisis prompts.
Chandra et al. surveyed 283 adults in the United States with lived experience of mental health challenges \cite{chandra2025lived}.
From their accounts, the authors built a taxonomy of 19 harmful AI behaviors and 21 negative psychological impacts.
Ajmani et al. collected accounts from 53 people who had used conversational AI during a mental health crisis and interviewed 16 mental health experts \cite{ajmani2026seeking}.
People turned to AI when they lacked access to professionals or feared burdening others.
The authors recommend designing such systems as a bridge to human support.
Arnaiz-Rodriguez et al. audited five LLMs on about 2{,}000 crisis messages \cite{arnaizrodriguez2026between}.
Replies to self-harm and suicidal ideation received the lowest scores.
The authors ran their experiments outside the United States and argue that generic or US-centric contact information undermines trust in such settings.
Sambasivan et al. made a related argument about algorithmic fairness research, which they found to be West-centric in its subgroups, values and methods \cite{sambasivan2021reimagining}.
Their interviews in India showed that several assumptions of this research do not hold there.

None of the conversational agents above was built for or tested with YACs.
The companion tested models with prompts or studied companion users, peer supporters and adults with lived experience of mental health challenges.
The two dementia tools serve family caregivers of people with dementia, and CareCall serves socially isolated people.
Shankar et al. is the only work above that focuses on young caregivers \cite{shankar2026effectiveness}.
Their review summarizes existing digital tools and does not build or evaluate an LLM system.
Carebot is designed for YACs caring for a loved one with any health condition.
Like Parmanto et al., we combined FT with retrieval in some configurations.
In our comparison, retrieval did not improve the replies, so the final Carebot uses FT alone.

\subsection{Existing Caregiver Support and Evaluation of AI Companions}\label{sec:rw-eval}

Existing support for young carers includes online resources, local services, and peer support. In the United Kingdom, Carers Trust supports young adult carers aged 16 to 25 through local organizations offering practical and emotional support, advice, and activities \cite{carerstrust2026about,carerstrust2026support}. The Children's Society, through its Include programme, provides information on education, employment, well-being, and carers' rights, along with resources for finding local young-carer services \cite{childrenssociety2026include}. The NHS also provides guidance on accessing support through schools, healthcare professionals, and local services \cite{nhs2024youngcarers}. Action for Children's Sidekick, initially developed for young carers, provides confidential text and email support from human coaches. It primarily serves young people aged 13 to 18, with some eligibility extending to age 25 \cite{actionforchildren2026sidekick}.

In India, Carers Worldwide India works with local organizations to provide peer support groups, counselling, livelihood assistance, and access to government entitlements for unpaid family caregivers \cite{carersworldwide2026india,carersworldwide2019impact}. Its programmes support family caregivers more broadly rather than focusing specifically on YACs \cite{carersworldwide2026india}. These services provide informational, practical, and interpersonal support, although their availability and target populations vary across countries.

\section{Methodology}\label{sec:method}
 
Our study had two parts.
In Part A, we built the carebot, compared four different configurations offline.
In Part B, we compared the base model with Carebot in two ways, through an metric evaluation by human experts and an LLM judge, and through a user study with YACs (Section~\ref{sec:study}).
 
\subsection{Part A. System Design}\label{sec:design}
 
Part A had five steps.
We selected a base model, built a training corpus, fine-tuned the model, and added an optional retrieval component.
We then compared the resulting configurations offline.
 
\subsubsection{Architecture}\label{sec:architecture}
 
\begin{figure}[t]
  \centering
  \includegraphics[width=0.75\linewidth]{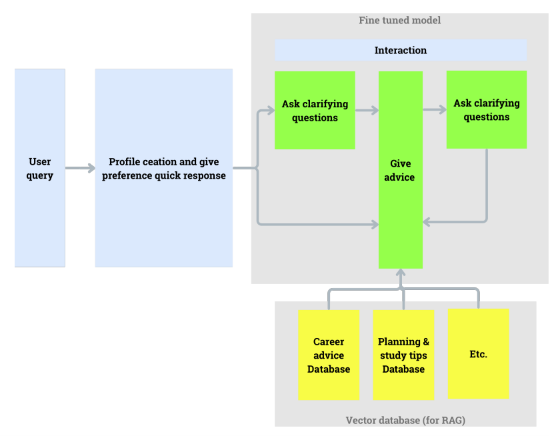}
  \caption{Overview of Carebot. The user's profile and preference for a quick reply shape the system prompt. The fine-tuned model either asks clarifying questions or gives advice. In the conditions with retrieval, a vector database supplies passages from career and caregiving documents.}
  \Description{A flow diagram. A user query passes through a profile step to the fine-tuned model. The model either asks clarifying questions or gives advice. A vector database with career and caregiving documents feeds the model in the conditions with retrieval.}
  \label{fig:architecture}
\end{figure}

Carebot is a chat application embedded in an online survey.
Before the chat begins, the user answers survey questions about themselves and their caregiving situation.
The questions are adapted from Dang's thesis \cite{dang2024young}.
They cover, for example, age, country, the relationship to the care recipient, the care recipient's condition and the time spent on care in Appendix~\ref{app:config}.
The answers form a caregiving profile, which the application inserts into the system prompt.
The system prompt is the fixed set of instructions that the application sends to the model with every conversation (Appendix~\ref{app:prompt}).
It asks the model to act as a warm, empathic and supportive assistant that focuses on practical, everyday help and keeps answers concise but caring.
The profile also records whether the user prefers a quick response.
If so, the model gives advice at once.
Otherwise, it first asks one or two clarifying questions, which the prompt calls scaffolding.
Each user message also carries a short instruction to keep the response concise unless the user asks for details.
In the conditions with retrieval, the system adds passages from a document store to the prompt (Section~\ref{sec:rag}).
The application runs on Streamlit, and the fine-tuned model is hosted on Fireworks AI.
Figure~\ref{fig:architecture} gives an overview of the design.
 
\subsubsection{Base Model Selection}\label{sec:selection}

We first compared six open models on Human and LLM Mental Health Conversations, a public dataset previously used by Bird et al. \cite{bird2024generative}. Only the fine-tuned Gemma 3-27B gave more empathetic and less formal replies while staying coherent (Table~\ref{tab:selection} in Appendix~\ref{app:offline}). 
We therefore built Carebot on Gemma-3-27B-IT, the instruction-tuned release of this model.

\subsubsection{Dataset Composition}\label{sec:dataset}
We generated a synthetic corpus of 10{,}000 conversations between a caregiver and a supportive companion using GLM 5.2.
The generation prompts used transcripts from real caregivers as few-shot examples, taken from the sixty hours of interview data in Dang's thesis \cite{dang2024young}.
The corpus is balanced across the ten caregiving categories in Dang's thesis \cite{dang2024young}, 20 countries, 23 conditions of the care recipient, and gender.
The conditions are the health problems of the person receiving care, such as dementia, cancer and substance use.
Two percent of the conversations show the companion referring a user in crisis to safety resources.
A crisis is an immediate risk to the safety of the user or the care recipient.
Examples include thoughts of suicide or self-harm, abuse at home and medical emergencies.
Appendix~\ref{app:offline} describes the deduplication and quality checks.
We split the corpus into 9{,}500 training examples and 500 held-out examples, with 50 per caregiving category and a reference reply for each. We also obtained 54 question and answer passages from real interviews with caregivers.
Together with the 500 held-out examples, they form an evaluation set of 554 inputs.
 
\subsubsection{Fine-Tuning}\label{sec:finetuning}

We fine-tuned Gemma-3-27B-IT with low-rank adaptation (LoRA) \cite{hu2022lora} on the 9{,}500 training conversations (Section~\ref{sec:dataset}).
LoRA trains a small set of added parameters and leaves the original weights unchanged.
The loss covered only the assistant's turns, so tokens from user messages and padding did not contribute to training.
Fine-tuning had to teach the model the warm and brief style of the training conversations without harming its fluency.
The model selection experiment showed this risk.
Most fine-tuned models in that experiment produced incoherent or at times inappropriate replies (Section~\ref{sec:selection}).
We therefore chose settings that keep the fine-tuned model close to Gemma-3-27B-IT.
We used a LoRA rank of 8, a single epoch and learning rate of $10^{-5}$ as hyperparameters.
The training loss measures how poorly the model predicts the assistant's turns in the training data.
It fell from 5.34 to 1.71 over training.
Appendix~\ref{app:config} lists the remaining settings.
 
\subsubsection{Retrieval Component}\label{sec:rag}
 
The retrieval store holds web documents on young caregiving and career guidance, embedded with the all-mpnet-base-v2 sentence encoder \cite{reimers2019sentence,song2020mpnet} in a FAISS index \cite{douze2024faiss}.
Several documents describe services in the United Kingdom and the Netherlands (Table~\ref{tab:ragsources} in Appendix~\ref{app:config}).
The first setup inserted three unlabeled passages into the prompt.
The model then treated other caregivers' interview statements as the user's own history, a context leak.
The revised setup labels each passage by its source and tells the model not to attribute retrieved content to the user.
It also gates retrieval and adds up to three passages only when their cosine similarity to the user message is at least 0.60.
This threshold is met for about 20\% of messages. We did not include retrieval in the final version of Carebot as most documents in the store come from particular countries.
Their information is often specific and may not apply to users elsewhere.
Retrieved passages may therefore be irrelevant to users in other countries.
The metric evaluation also found that retrieval did not improve replies on average (Section~\ref{sec:results-variants}).

\begin{table}[t]
  \caption{The four conditions in the metric evaluation.}
  \label{tab:conditions}
  \begin{tabular}{lll}
    \toprule
    Condition & Model & Retrieval \\
    \midrule
    Base & Gemma-3-27B-IT & None \\
    RAG & Gemma-3-27B-IT & Labeled, not gated \\
    FT (Carebot) & Fine-tuned & None \\
    FT + gated RAG & Fine-tuned & Labeled, gated \\
    \bottomrule
  \end{tabular}
\end{table}

\subsubsection{Evaluated Conditions}\label{sec:variants}
 
Table~\ref{tab:conditions} lists the four evaluated conditions: Base, Base + RAG, FT and FT + RAG. Based on the results presented in Section~\ref{sec:results-variants}, Carebot is the FT condition, the fine-tuned model without retrieval, and the user study used this configuration. 
 
\subsubsection{Metric Evaluation}\label{sec:expert}

We gathered three tiers of evidence.
Tier one used reference-based metrics, tier two used LLM judges, and tier three used human review by the research team.
The three tiers give scores or counts.
Human review gives counts of preferred replies and written notes on the reasons for each preference.
We treated human review as the primary tier, so its verdicts took precedence when the tiers disagreed.

\begin{itemize}
  \item \textbf{Reference-based metrics.}
  The first tier compared each generated reply with the reference reply for the same input on the 500 synthetic held-out examples.  Table~\ref{tab:metrics} in Appendix~\ref{app:offline} describes each metric.
  BLEU \cite{papineni2002bleu}, ROUGE-L \cite{lin2004rouge} and METEOR \cite{banerjee2005meteor} measure overlap in wording.
  BERTScore \cite{zhang2020bertscore} and BLEURT-20 \cite{sellam2020bleurt} measure similarity in meaning, so they also give credit for paraphrases.
  These metrics show how closely a reply resembles the reference reply.
  They do not measure whether users find the reply helpful.
  The reference replies also come from the same synthetic corpus as the training data.
  The metrics therefore reward replies that match the style of that corpus.

  \item \textbf{LLM judges.}
  In LLM-as-a-judge evaluation, an LLM scores or compares replies in place of human raters \cite{zheng2023judging}.
  We used GLM 5.2 and DeepSeek-V4-Pro as judges on the 54 real inputs, which have no reference replies.We chose judges from model families other than Gemma, which underlies every condition, because LLM judges can favor their own outputs \cite{panickssery2024llm}.
  DeepSeek-V4-Pro served as an independent check on GLM 5.2, because GLM 5.2 also generated the training corpus.
  In tier two, each judge compared the replies of every condition.
  We kept a verdict only when the judge chose the same reply in both orders.
  This protocol controls for position bias, the tendency of a judge to favor the reply shown first.
  The judging prompt also stated that length is not a merit.

  \item \textbf{Human expert review.}
  In tier three, the human expert team reviewed samples of replies that the team had selected at random.
  The reviewers did not know which condition had produced each reply.
  Each of the three rounds used 15 items per condition.
  Round one compared the fine-tuned version against the base model.
  Round two examined fine-tuned against fine-tuned with retrieval after we fixed the context leak (Section~\ref{sec:rag}).
  Round three compared fine-tuned against fine-tuned with retrieval again after gating the relevant context.
  In each round, reviewers chose the reply they preferred and wrote notes on their reasons.
\end{itemize}

\begin{table}[h]
  \caption{Participants in the analysis set by condition. Counts show the number of participants, with percentages of the condition in parentheses.}
  \label{tab:participants}
  \begin{tabular}{lcc}
    \toprule
    & Base ($n = 45$) & Carebot ($n = 45$) \\
    \midrule
    \multicolumn{3}{l}{\textit{Age}} \\
    Mean age in years & 19.5 & 18.7 \\
    Median age in years & 19 & 18 \\
    Age range in years & 17 to 23 & 17 to 24 \\
    \midrule
    \multicolumn{3}{l}{\textit{Occupation}} \\
    Student & 35 (78\%) & 43 (96\%) \\
    Student and working part-time & 4 (9\%) & 1 (2\%) \\
    Working professional & 6 (13\%) & 1 (2\%) \\
    \midrule
    \multicolumn{3}{l}{\textit{Use of AI chatbots}} \\
    Daily & 22 (49\%) & 26 (58\%) \\
    A few times a week & 19 (42\%) & 16 (36\%) \\
    Rarely & 3 (7\%) & 3 (7\%) \\
    Never or first time & 1 (2\%) & 0 (0\%) \\
    \bottomrule
  \end{tabular}
\end{table}

\subsection{Part B. User Study Design}\label{sec:study}
 
\subsubsection{Participants and Recruitment}\label{sec:participants}

We evaluated Carebot with 90 YACs, 45 who interacted with the base model and 45 with Carebot.
Most were students enrolled in a university (87\%).
They most often cared for a grandparent (24\%), a parent (10\%), a sibling (9\%) or a friend (9\%), and several named more than one person. Table~\ref{tab:participants} describes the participants in each condition.
Participants in both groups had a mean age below 20, and most used AI chatbots daily or a few times a week.
The Carebot group had more students, at 96\% compared with 78\% in the base group.

The study compared two models.
In the base model, participants interacted with Gemma-3-27B without FT or RAG.
In the Carebot condition, participants interacted with the fine-tuned only version of Gemma-3-27B.

\subsubsection{Procedure}\label{sec:procedure}

Each participant used the Carebot web application, which connected them to the model.
Before the chat, participants filled in a form about their caregiving situation, including the health condition of the person they care for.
The application passed these answers to the model so that it could personalize its replies.
They then interacted with the model about topics of their own choice.
After the conversation, they completed an online questionnaire about their experience, which Appendix~\ref{app:survey} lists in full.
The application did not store the conversations, so all user study results in this paper come from the questionnaire.

\subsubsection{Measures}\label{sec:measures}

The first part of the questionnaire described the participant.
It asked for the participant's age, occupation, frequency of AI chatbot use and caregiving status.
The caregiving item asked whether the participant currently helped care for someone who was ill, disabled, elderly or living with a health or mental condition.
We counted participants who reported any level of help with care as having caregiving duties (Table~\ref{tab:survey} in Appendix~\ref{app:survey}).
Conversation items covered an overall rating (1 to 10), a tone rating (1 to 5), response length, the balance between emotional support and practical advice, whether a response felt particularly right or wrong, and a comparison with general AI tools such as ChatGPT or Claude.

\subsubsection{Statistical Analysis}\label{sec:analysis}

We compared the base and Carebot conditions on nine outcomes from the questionnaire described in Section~\ref{sec:measures}.
Two of them are ratings, namely the overall rating from 1 to 10 and the tone rating from 1 to 5.
The other seven are yes or no outcomes, which record whether a participant chose a given answer, such as finding the replies too practical (Table~\ref{tab:survey}).
For ratings, we used the Mann-Whitney U test, which checks whether one group tends to rate higher without assuming normally distributed ratings \cite{mann1947test}.
For yes or no outcomes, we used Fisher's exact test, which compares how often each group gave an answer and stays valid in small samples \cite{fisher1922interpretation}.
For yes or no outcomes, we report conditional maximum likelihood odds ratios (ORs) with exact 95\% confidence intervals (CIs) \cite{agresti2013categorical}.
We adjusted the p-values of the nine outcomes in Table~\ref{tab:outcomes} with the Holm correction, which limits chance findings when many outcomes are tested \cite{holm1979simple}.
We used a significance level of 0.05.

\section{Results}\label{sec:results}
 
We first report the metric comparison of the four conditions, including Carebot against the base model.
We then report the user study.
 
\subsection{Part A. Metric Comparison}\label{sec:results-variants}

This section reports how the four conditions in Table~\ref{tab:conditions} performed across the three tiers of the metric evaluation (Section~\ref{sec:expert}).
We report the reference-based metrics, the pairwise verdicts of the LLM judges, the problems that retrieval raised, the human expert review and the final selection.
 
\paragraph{Reference-based metrics}
Table~\ref{tab:refmetrics} reports the five metrics for the four conditions.
Fine-tuning brought the replies closer to the reference replies.
The FT condition scored highest of the four conditions on BERTScore, BLEU and ROUGE-L.
Its BERTScore was 0.5756 against 0.5346 for the base model, and its ROUGE-L was 0.1950 against 0.1570.
BLEU scores stayed below 0.014 for every condition, because a free reply rarely repeats the exact word sequences of a reference.
METEOR and BLEURT-20 favored the base model, with 0.2954 against 0.2577 for METEOR and 0.4735 against 0.4470 for BLEURT-20 .
A likely reason is that the longer base replies cover more of the content of the reference replies.
Retrieval made almost no difference.
Adding retrieval changed each metric by less than 0.012 for the base model and by less than 0.005 for the fine-tuned model.
The reference replies come from the same synthetic corpus as the training data.
These gains therefore suggest that fine-tuned learned the wording and style of that corpus.
They do not show whether users prefer its replies, which the user study tests (Section~\ref{sec:results-study}).

\paragraph{Pairwise judging}
We ran two pairwise comparisons (Section~\ref{sec:expert}), the base model against the fine-tuned model and the fine-tuned model against the fine-tuned model with gated retrieval.
Table~\ref{tab:variants} reports the results.
GLM 5.2, the primary judge, preferred the fine-tuned model to the base model in 39 of the 46 pairs it counted (85\%).
It discarded 8 of the 54 pairs.
Retrieval added passages to the prompt for only 28 of the 54 inputs, so the comparison with retrieval used these 28 inputs.
On them, GLM 5.2 preferred the fine-tuned model without retrieval in 14 of 23 counted pairs (61\%) and discarded 5.
DeepSeek-V4-Pro, the secondary judge, preferred each model equally often, 14 times each, and discarded 26 of 54 pairs.
In the comparison with retrieval, it preferred the fine-tuned model without retrieval in 9 of 15 counted pairs (60\%) and discarded 13.

The two judges differed in position bias, which is a tendency to prefer the reply shown first whatever its content.
An unbiased judge would choose the first reply about half of the time.
Counting each order separately, GLM 5.2 chose the first reply 50\% of the time in the comparison with retrieval, which shows no measurable bias.
Counting the same way, DeepSeek-V4-Pro chose the first reply 70\% of the time against the base model.
A judge with this bias often picks a different reply when the order is swapped, so the pair is discarded.
This explains why DeepSeek-V4-Pro discarded far more pairs than GLM 5.2.

\begin{table}[t]
  \caption{Reference-based metrics for the four conditions on the 500 synthetic held-out examples. Higher scores mean greater similarity to the reference replies. Bold marks the highest score for each metric.}
  \label{tab:refmetrics}
  \begin{tabular}{lccccc}
    \toprule
    Condition & BERTScore & BLEU & ROUGE-L & METEOR & BLEURT-20 \\
    \midrule
    Base & 0.5346 & 0.0075 & 0.1570 & \textbf{0.2954} & \textbf{0.4735} \\
    RAG & 0.5440 & 0.0067 & 0.1582 & 0.2842 & 0.4652 \\
    FT (Carebot) & \textbf{0.5756} & \textbf{0.0135} & \textbf{0.1950} & 0.2577 & 0.4470 \\
    FT + RAG & 0.5746 & 0.0123 & 0.1942 & 0.2535 & 0.4451 \\
    \bottomrule
  \end{tabular}
\end{table}
 
\begin{table}[t]
  \caption{Pairwise judging on the 54 real inputs. Each row compares condition A (named first) with condition B (named second). For each input, the judge read one reply from A and one from B and picked the better one. The judge saw each pair twice, with the order swapped the second time. A condition wins when the judge picked its reply both times. If the judge picked differently the two times, the pair is discarded. The win share of B is the percentage of wins that went to B. For example, GLM 5.2 picked FT over Base 39 times and Base over FT 7 times, so the win share of FT is 39 of 46, or 85\%. The last two rows use only the 28 inputs for which retrieval added passages.}
  \label{tab:variants}
  \begin{tabular}{llcccc}
    \toprule
    Comparison (A against B) & Judge & A wins & B wins & Discarded & Win share of B \\
    \midrule
    Base against FT (Carebot) & GLM 5.2 & 7 & 39 & 8 & 85\% \\
    Base against FT (Carebot) & DeepSeek-V4-Pro & 14 & 14 & 26 & 50\% \\
    FT against FT + RAG & GLM 5.2 & 14 & 9 & 5 & 39\% \\
    FT against FT + RAG & DeepSeek-V4-Pro & 9 & 6 & 13 & 40\% \\
    \bottomrule
  \end{tabular}
\end{table}
 
\paragraph{Retrieval}
Retrieval raised two problems during evaluation.
First, the initial setup caused context leaks, in which the model treated other caregivers' interview statements as the user's own history.
Labeling each passage by its source removed the leaks we observed.
Second, gated retrieval at first appeared to raise the single-response score of the fine-tuned model by 0.16 points on the 1 to 5 scale.
The 54 real inputs were also put in the retrieval store, so retrieval often returned the user's own message.
After we excluded these self-matches, the gain shrank to 0.04 points (Appendix~\ref{app:offline}).
The gate fired on 28 of the 54 real inputs.
As the pairwise results above show, both judges preferred the fine-tuned model without retrieval on these inputs (Table~\ref{tab:variants}).
 
\paragraph{Human review}
Human expert team reviewed samples of replies in three rounds.
Round one compared the fine-tuned against the base model.
Reviewers clearly preferred the fine-tuned replies.
They described the base replies as good but generic and the fine-tuned replies as warmer and less templated.
Round two examined retrieval after the context-leak fix.
Reviewers found that retrieval helped only when it surfaced relevant information and degraded replies when it added unrelated or generic content.
Round three compared the fine-tuned model with and without gated retrieval.
Reviewers preferred the fine-tuned model without retrieval on 11 of 15 items and with retrieval on 2, with 2 ties.
Human review therefore agreed with both debiased judges on retrieval.
 
\paragraph{Selection}
We selected the fine-tuned model without retrieval for Carebot.
It scored highest on BERTScore and ROUGE-L, and GLM 5.2 preferred it to the base model.
Both judges and the human experts also preferred it to its variant with gated retrieval.
 
\subsection{Part B. User Study}\label{sec:results-study}
 
\subsubsection{Participants}
 
Table~\ref{tab:participants} describes the 90 participants in the analysis set, 45 per condition.
All participants were aged 17 to 24.
The two groups differed in composition.
About seven in ten base participants reported caregiving duties, compared with about half of Carebot participants.
 
\subsubsection{Overall Ratings}
 
After the conversation, participants gave Carebot and the base model an overall rating on a scale from 1 to 10, where higher scores mean a more favorable rating.
This single item captures how positively participants judged the companion as a whole.
The mean overall rating was 7.44 (median 8) for the base model and 7.38 (median 7) for Carebot, so participants in both conditions rated their companion fairly positively and almost identically.
We compared the two groups with the Mann-Whitney U test, which checks whether ratings in one group tend to be higher than in the other without assuming that ratings follow a bell-shaped distribution.
The difference was not significant ($p = 0.30$), which means that a gap this small could easily arise by chance.
Low ratings of 5 or below were also similar, coming from 6 of 45 base participants (13\%) and 4 of 45 Carebot participants (9\%) ($p = 0.74$).
 
\subsubsection{Response Balance}\label{sec:balance}
 
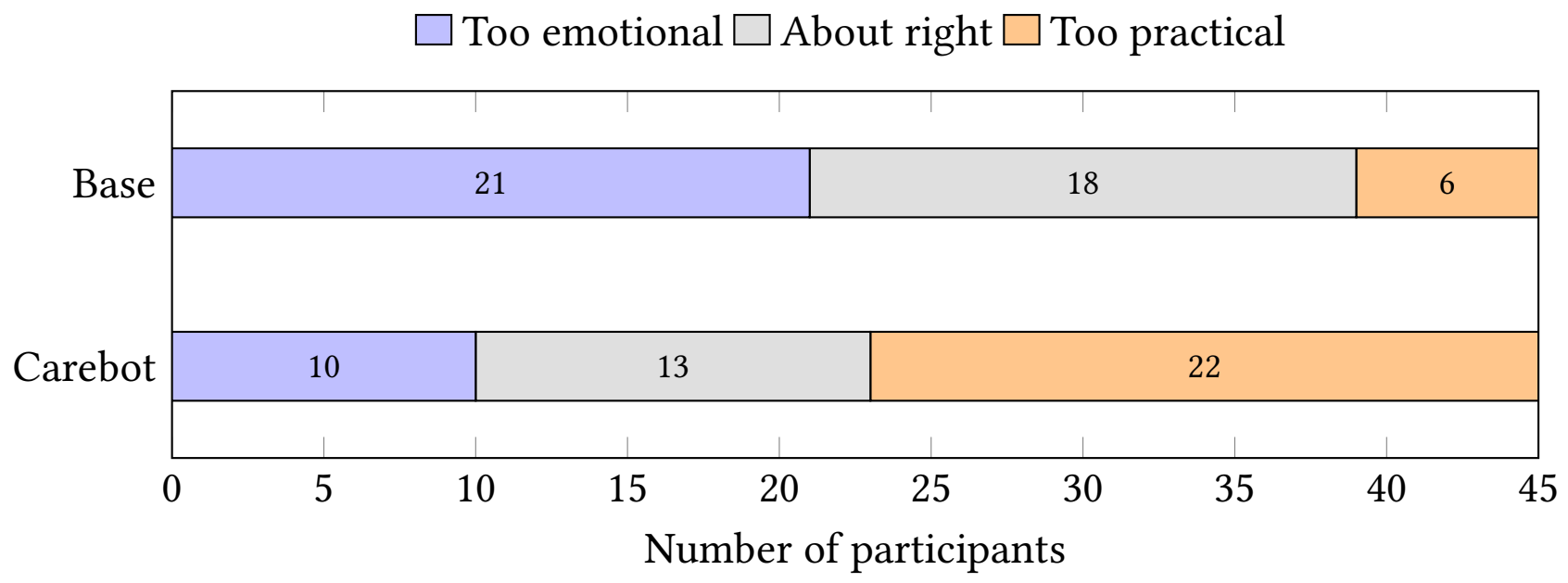
\begin{figure}[t]
  \centering
  \begin{tikzpicture}
    \begin{axis}[
      xbar stacked,
      width=0.75\linewidth,
      height=4.2cm,
      bar width=14pt,
      xmin=0, xmax=45,
      xlabel={Number of participants},
      symbolic y coords={Carebot,Base},
      ytick=data,
      enlarge y limits=0.5,
      legend style={at={(0.5,1.05)}, anchor=south, legend columns=3, draw=none},
      nodes near coords,
      every node near coord/.append style={font=\footnotesize, color=black},
    ]
      \addplot[fill=blue!25] coordinates {(21,Base) (10,Carebot)};
      \addplot[fill=gray!25] coordinates {(18,Base) (13,Carebot)};
      \addplot[fill=orange!45] coordinates {(6,Base) (22,Carebot)};
      \legend{Too emotional, About right, Too practical}
    \end{axis}
  \end{tikzpicture}
  \caption{Answers to the response balance item by condition, with 45 participants per condition. Participants who used Carebot chose ``too practical'' nearly four times as often as participants who used the base model. Base participants more often chose ``too emotional''.}
  \Description{Two horizontal stacked bars. The base bar has 21 too emotional, 18 about right and 6 too practical. The Carebot bar has 10 too emotional, 13 about right and 22 too practical.}
  \label{fig:balance}
\end{figure}

The clearest difference concerned the balance between emotional and practical content.
The questionnaire asked whether the replies gave the right balance of emotional support and practical advice.
Figure~\ref{fig:balance} shows the answers.
About half of Carebot participants found the replies too practical (22 of 45, 49\%), compared with about one in eight base participants (6 of 45, 13\%).
This difference was significant (Fisher's exact test, $p < 0.001$) and remained significant after the Holm correction (adjusted $p = 0.005$).

Base participants more often chose ``too emotional'' (21 of 45, 47\%) than Carebot participants (10 of 45, 22\%).
This difference is significant only before correction ($p = 0.026$), and the outcome was not part of the corrected family of nine tests, so we interpret it with caution.
The share choosing ``about right'' did not differ significantly (18 of 45, 40\%, for base compared with 13 of 45, 29\%, for Carebot; $p = 0.38$).
 
\subsubsection{Tone and Descriptions of Replies}
 
Table~\ref{tab:outcomes} lists all nine outcomes.
Apart from \hyperref[sec:balance]{\textit{response balance}}, no difference remained significant after the Holm correction (all adjusted $p \geq 0.21$), so we treat the following results as exploratory.
Base participants rated the tone higher, with a mean of 4.09 compared with 3.73 on a scale from 1 to 5 ($p = 0.026$, adjusted $p = 0.21$).
They more often described the replies as extremely polite and helpful, with 17 of 45 (38\%) compared with 7 of 45 (16\%) ($p = 0.031$).
Carebot participants more often said that nothing in the replies stood out.
This was the case for 21 of 45 Carebot participants (47\%) and 11 of 45 base participants (24\%) ($p = 0.047$).
Carebot participants also more often found the replies too short or lacking depth (18 of 45 against 9 of 45, $p = 0.065$).
Some participants had used general-purpose assistants.
Among them, 15 of 40 base participants (38\%) rated the companion as better, compared with 10 of 39 Carebot participants (26\%) ($p = 0.33$).
 
\begin{table*}[t]
  \caption{User study outcomes in the analysis set. Ratings show means, with medians in parentheses for the overall rating. Yes or no outcomes show counts and percentages. ORs above 1 mean the outcome was more common with Carebot. Holm-adjusted p-values cover the nine outcomes in this table.}
  \label{tab:outcomes}
  \begin{tabular}{lccccc}
    \toprule
    Outcome & Base & Carebot & OR (95\% CI) & $p$ & Holm $p$ \\
    \midrule
    Overall rating, 1 to 10 & 7.44 (8) & 7.38 (7) & n/a & 0.30 & 0.89 \\
    Tone rating, 1 to 5 & 4.09 & 3.73 & n/a & 0.026 & 0.21 \\
        \textbf{Replies too practical} & \textbf{6/45 (13\%)} & \textbf{22/45 (49\%)} & \textbf{6.08 (2.02, 21.15)} & \textbf{{\boldmath$<$}0.001} & \textbf{0.005} \\
    Extremely polite and helpful & 17/45 (38\%) & 7/45 (16\%) & 0.31 (0.09, 0.91) & 0.031 & 0.21 \\
    Nothing stood out & 11/45 (24\%) & 21/45 (47\%) & 2.67 (1.01, 7.39) & 0.047 & 0.28 \\
    Too short or lacked depth & 9/45 (20\%) & 18/45 (40\%) & 2.64 (0.95, 7.78) & 0.065 & 0.32 \\
    A reply felt right & 26/45 (58\%) & 18/45 (40\%) & 0.49 (0.19, 1.22) & 0.14 & 0.56 \\
    Better than ChatGPT or Claude & 15/40 (38\%) & 10/39 (26\%) & 0.58 (0.19, 1.66) & 0.33 & 0.89 \\
    Overall rating of 5 or below & 6/45 (13\%) & 4/45 (9\%) & 0.64 (0.12, 2.92) & 0.74 & 0.89 \\
    \bottomrule
  \end{tabular}
\end{table*}

\subsubsection{Summary of User Study Results}

Participants rated the two companions similarly overall, so the user study did not show that fine-tuning improved the experience of YACs.
The one difference that remained significant after correction for multiple comparisons was \hyperref[sec:balance]{\textit{response balance}}: Carebot participants were significantly more likely to describe its replies as too practical.
The base model's higher tone and politeness ratings did not remain significant after correction, and the Carebot participants' tendency to find replies short or lacking depth was not significant.

\section{Discussion}\label{sec:discussion}
This section interprets the main result of the study. The metric evaluation favored Carebot, yet participants rated it no higher overall and more often found its replies too practical.
We examine why the two evaluations diverged.

\subsection{Why Metric Evaluation and User Ratings Differed}\label{sec:disc-eval}

First, a system can reply well to one isolated message and still fall short in a conversation, where each reply must build on earlier ones.
Lee et al. found that better performance without interaction does not always lead to better interaction with people \cite{lee2023evaluating}.
In our metric evaluation, the offline methods scored single replies to single messages, while participants judged whole conversations about their own situations.

Second, LLM judges can favor text written in their own style \cite{panickssery2024llm}.
In our metric evaluation, GLM 5.2 wrote the training corpus, the reference replies came from that corpus, and GLM 5.2 also served as the primary judge.
The reference metrics and the primary judge therefore partly measured agreement with GLM 5.2.
DeepSeek-V4-Pro had no part in the training data, and it split evenly between Carebot and the base model, although its strong position bias makes this split hard to interpret.

Third, observers and recipients can judge empathy differently.
Clinicians who read patient questions rated chatbot answers as more empathetic than physician answers \cite{ayers2023comparing}.
People who took part in conversations rated chatbots as less empathetic than human partners \cite{liu2025illusion}.
In our metric evaluation, the human reviewers and the LLM judges evaluated replies to queries that YACs had written. However, the YACs involved in the user study rated replies to their own concerns.

Fourth, a model tested on data that resembles its training data can appear stronger than it is in real use \cite{hutchinson2022evaluation}.
In our metric evaluation, the offline inputs were synthetic messages and passages from caregiver interviews, while the YACs in the user study raised their own concerns.
A model tuned and tested on the offline inputs may therefore fit poorly with what YACs asked.

\subsection{What the Metric Evaluation Missed}\label{sec:disc-missed}

The metric evaluation did not measure three aspects that could have mattered to YACs or to their safety.

Firstly, it did not measure the balance between emotional support and practical advice.
A reply can be warm and give useful advice, and still fail a user who only wanted someone to listen.
In the user study, the question on this balance showed the clearest difference between Carebot and the base model, yet none of the measures in our metric evaluation asked about it.
This gap matters for many YACs, who said in Dang's interviews that they wanted professionals who listen without judging them \cite{dang2024exploring}.

Secondly, it did not capture whether the replies were long and substantial enough.
Fine-tuning taught Carebot the brief style of the training corpus.
LLM judges tend to prefer longer replies whatever their quality, a tendency known as length bias \cite{zheng2023judging}, so the pairwise prompt told the judges not to rate a reply higher only because it was longer.
This removed a known bias, but it may also have hidden a signal that participants used.
Carebot participants more often chose the answer ``too short or lacked depth'' (18 of 45 against 9 of 45), although this difference was not statistically significant (Table~\ref{tab:outcomes}).
That answer combines brevity with a lack of substance, such as little engagement with the participant's situation, and because we did not store conversations, we cannot tell which of the two participants meant.

Thirdly, it did not test whether crisis responses were locally relevant.
A participant reported that the base model answered a disclosure of suicidal thoughts with helplines for the United States and Canada.
No part of our metric evaluation tested whether crisis referrals point to services in the user's country.
These gaps share a common source: YACs had no role in the metric evaluation.

\subsection{Alternative Interpretations}\label{sec:disc-alt}

The results also allow two other interpretations, which concern the system prompt and the choice of configuration.

First, the system prompt may explain part of the shift.
Both conditions used the same prompt, which asked for practical, everyday help and concise answers (Appendix~\ref{app:prompt}).
The fine-tuned model, trained on brief replies, may have followed these instructions more closely than the base model.
Varying the prompt while holding the model fixed would separate the effect of the prompt from the effect of fine-tuning.

Second, a different configuration would likely not have changed the main result.
Retrieval added passages to only about one in five user messages (Section~\ref{sec:rag}), and most of those passages gave practical guidance on careers and services (Table~\ref{tab:ragsources}).
Adding retrieval to Carebot would therefore have left most replies unchanged and made the rest more practical.
Retrieval without fine-tuning would have kept the style of the base model, which nearly half of its users found too emotional.
We did not test these configurations with users, but reviewers preferred Carebot without retrieval on 11 of 15 items (Section~\ref{sec:results-variants}).

\subsection{Limitations}\label{sec:limitations}
Majority of our YACs were male university students.
We also assigned conditions by time slot, so the two groups differed in caregiving status and age.
A future study should assign equal YACs aged 18 to 25 to conditions at random.
It could also test a version of Carebot that first asks whether the user wants to be heard or wants advice. The application did not store conversations, so we cannot link ratings to specific replies or verify the reported crisis response.
A longer deployment with consented logging would address this.
GLM 5.2 wrote the training corpus and reference replies and also served as the primary judge, and the second judge showed a strong position bias.
No YACs took part in the metric evaluation.
Future work could fine-tune on replies that YACs write or rate, and test which LLM judges agree with YAC ratings.
Finally, Carebot has no source of local information, and we did not test whether crisis referrals point to services in the user's country.
Retrieval from local sources, such as services and helplines in India, could improve both.
 
\section{Conclusion}\label{sec:conclusion}
We presented Carebot, a companion for young adult caregivers built on a fine-tuned Gemma-3-27B-IT, and compared its metric evaluation with how it fared with the people it was built for. On paper, Carebot did well: three well-being experts chose the fine-tuned version over RAG and the combined approach, and an LLM judge preferred its replies to the base model’s in 85\% of pairs. However, the 90 YACs in our user study saw it differently. Many found Carebot’s replies short and too practical, offering advice when they wanted to feel heard, while the base model came across as more emotionally attuned. Expert review and automated judges are useful, but they cannot stand in for the people a system is meant to help. Teams building support companions for YACs should involve YACs in the evaluation itself, letting them define what a good reply looks like and when practical advice is welcome. 
 
\section*{Ethical Considerations Statement}
The young people in this study care for family members, often while also studying, working, or simply growing up, and many of them are under real strain. That shaped how we handled both the people and their words.

The study was approved by the Institutional Review Board of University of Amsterdam (ref. 2025-898). The 54 interview passages we use for retrieval and evaluation come from an existing dataset collected under \citet{dang2024young}. Everyone we interviewed was 18–25 years old and gave written informed consent, which covered the reuse of anonymised excerpts in research of this kind. They knew they could skip questions, stop at any point, or withdraw their data up to October 01, 2026.

Before using any passage, we removed names, places, dates, and anything about the family member’s condition that could identify them. The people being cared for never agreed to take part, so we were just as careful with their details as with the participants’ own. The passages stay on our secure institutional storage and only the research team can access them. We aren’t releasing them. Instead, we will share code, prompts, rubrics, and synthetic examples used in this study.

We are also wary of what a system like this could do. A companion that sounds warm can still give bad advice about medication or benefits, miss signs that someone is in crisis, or become a substitute for people who could actually help. Our prototype hasn’t been used by caregivers outside the study. One of our main findings is that standard metrics can score a response well while missing exactly these problems. A good score shouldn’t be read as a sign that the system is safe, and anyone building on this work should involve young carers and the organisations that support them from the start.
 
\section*{Generative AI Usage Statement}
We acknowledge the use of AI tools like Claude and ChatGPT for assistance
in coding and in improving the grammatical precision and presentation of this paper.
Within the study itself, GLM 5.2 generated the synthetic training corpus (Section~\ref{sec:dataset}).
GLM 5.2 and DeepSeek-V4-Pro served as the LLM judges described in Section~\ref{sec:expert}. The experimental design, analysis, and all scientific conclusions accurately represent the original contributions of the
authors.
 
\bibliographystyle{ACM-Reference-Format}
\bibliography{references}
 
\appendix
 
\section{Model Configuration}\label{app:config}
 
Fine-tuning used the Unsloth library with the PEFT and SFTTrainer libraries from Hugging Face.
The fine-tuned model used LoRA with rank 8, one epoch and a learning rate of $10^{-5}$.
The 27B models were trained on NVIDIA A100 (40 GB) or H100 (94 GB) GPUs.
Smaller models and test runs used a GPU with 24 GB of memory.
Evaluation runs served the models with vLLM, and the fine-tuned model for the application is hosted on Fireworks AI.
 
Retrieval used chunks of 1{,}000 characters with an overlap of 100 characters, embedded with all-mpnet-base-v2 and stored in a FAISS index.
The embeddings are normalized, and the same encoder embeds documents and queries.
The first setup inserted the top three passages from maximal marginal relevance search.
The gated setup inserts up to three passages whose cosine similarity to the query is at least 0.60.
Table~\ref{tab:ragsources} lists the documents in the first store.
 
\section{System Prompt}\label{app:prompt}

The application sent the main instruction prompt below to the model with every conversation.
It filled the field \texttt{\char`\{user\_profile\char`\}} with the user's caregiving profile and the field \texttt{\char`\{context\char`\}} with the retrieved passages.
Each user query also carried the additional instruction in the second box.

\begin{tcolorbox}[colback=gray!5, colframe=gray!70, coltitle=white, fonttitle=\bfseries, title={Main instruction prompt}]
\ttfamily\small
You are Carebot, a warm, empathic, supportive assistant. You speak kindly and clearly, avoid medical or legal claims, and focus on practical, everyday help. Reflect feelings, validate, suggest small steps and resources, and encourage seeking trusted adults or professionals when appropriate. Keep answers concise but caring. Use the user's profile to personalize your tone and suggestions:\\
\char`\{user\_profile\char`\}\\
If the user prefers a quick response, immediately give your advice. If not, employ scaffolding techniques and ask one or two questions to better understand the situation before giving advice. Keep this scaffolding process concise.\\
You may use the retrieved context to help you answer the question, if relevant:\\
\char`\{context\char`\}\\
Here is the user query:
\end{tcolorbox}

\begin{tcolorbox}[colback=gray!5, colframe=gray!70, coltitle=white, fonttitle=\bfseries, title={Additional instruction prompt for each user query}]
\ttfamily\small
Keep your response concise, except when asked for details
\end{tcolorbox}
 
Each user query also carries the instruction ``Keep your response concise, except when asked for details''.
 
\begin{table}[h]
  \caption{Documents in the first retrieval store. The second store caps the first document at 300 chunks and adds five caregiver-specific sources.}
  \label{tab:ragsources}
  \small
  \begin{tabular}{p{0.55\linewidth}l}
    \toprule
    Title & Source \\
    \midrule
    Find a fulfilling career that does good & 80{,}000 Hours \\
    Framework for career guidance for young people & Euroguidance \\
    Professionalism of career guidance & Euroguidance \\
    Young carers: Shaping our future & The Children's Society \\
    Balancing caregiving and career: tips for starting your new job strong & Bright Horizons \\
    Help for young carers & NHS \\
    Do you know your rights in youth care? & Jeugdstem \\
    \bottomrule
  \end{tabular}
\end{table}
 
\section{Survey Instrument}\label{app:survey}
 
Table~\ref{tab:survey} lists the questionnaire items with the answer options observed in the two conditions.
Items marked as free text had no fixed options.
The questionnaire also asked for consent, the participant's full name and an email address.
 
\begin{table}[h]
  \caption{Questionnaire items and answer options.}
  \label{tab:survey}
  \small
  \begin{tabular}{p{0.42\linewidth}p{0.5\linewidth}}
    \toprule
    Item & Answer options \\
    \midrule
    Age & Number \\
    Gender & Male, female \\
    Current primary occupation or status & Student, working professional, student and working part-time \\
    Do you currently help look after a family member or loved one who is ill, disabled, elderly, or struggling with health or mental conditions? & Yes, a lot (high involvement caregiver), yes, a little, or left blank \\
    If yes, who are you helping care for? & Free text \\
    How often do you use AI chatbots (e.g. ChatGPT, Claude, Gemini)? & Daily, a few times a week, rarely, never or first time \\
    What specific topics or questions did you ask the AI model? & Free text \\
    Overall rating of the AI companion & 1 to 10 \\
    Tone, empathy, and accuracy & Extremely, very, moderately, slightly, or not at all polite or helpful \\
    Did Carebot give the right balance of emotional support and practical advice? & Far too emotional, not practical enough, somewhat emotional, not practical enough, about right, somewhat practical, not emotional enough, far too practical, not emotional enough \\
    Was there one response that felt particularly right or wrong? & Yes, it felt right, yes, it felt wrong, no, nothing stood out, with optional free text \\
    Response length and clarity & Just right, too short or lacked depth, too long or wordy \\
    How does this model compare to general AI tools like ChatGPT or Claude? & Much better, better, similar, worse, much worse, I haven't used them \\
    Comparison to a human companion & Free text \\
    What changes or tweaks would make this AI better for young adult caregivers? & Free text \\
    \bottomrule
  \end{tabular}
\end{table}
 \section{Sensitivity Analysis}\label{app:sensitivity}
 Table~\ref{tab:fullsample} reports the analysis of all responses received. P-values carry the Holm correction across the outcomes in the table.
The exception is the item on short replies, for which the table gives the uncorrected p-value from Fisher's exact test.
The balance p-value refers to the item as a whole.
 
\begin{table}[h]
  \caption{Outcomes of the base and Carebot participants. Ratings show means, with the median in parentheses for the overall rating.}
  \label{tab:fullsample}
  \begin{tabular}{lccc}
    \toprule
    Outcome & Base & Carebot & $p$ \\
    \midrule
    Overall rating, 1 to 10 & 7.38 (8) & 7.62 (7) & 0.95 \\
    Tone rating, 1 to 5 & 4.02 & 3.79 & 0.49 \\
    Balance, too emotional & 36\% & 18\% & \\
    Balance, about right & 47\% & 26\% & \\
    Balance, too practical & 17\% & 56\% & $< 0.001$ \\
    Too short or lacked depth & 25\% & 38\% & 0.17 \\
    Comparison with ChatGPT or Claude, 1 to 5 & 3.23 & 2.89 & 0.49 \\
    \bottomrule
  \end{tabular}
\end{table}
 
In covariate-adjusted ordinal models, the OR for response balance was 11.7 (95\% CI 4.3 to 31.8).
The OR for the overall rating was 2.0 ($p = 0.12$).
 
\section{Details of Part A}\label{app:offline}

\begin{table}[h]

  \caption{Reference-based metrics in the first tier. Each metric compares a generated reply with the reference reply for the same input. Higher scores mean greater similarity.}
  \label{tab:metrics}
  \small
  \begin{tabular}{lp{0.2\linewidth}p{0.52\linewidth}}
    \toprule
    Metric & Type & What it measures \\
    \midrule
    BLEU & Wording overlap & Share of word sequences of up to four words in the reply that also occur in the reference, with a penalty for short replies \\
    ROUGE-L & Wording overlap & Length of the longest sequence of words that occurs in the same order in both texts, with gaps allowed \\
    METEOR & Wording overlap & Matches of single words, including word stems and synonyms, with a penalty when matched words appear in a different order \\
    BERTScore & Meaning similarity & Similarity of the words in the two texts, based on representations from the language model BERT that depend on the surrounding words \\
    BLEURT-20 & Meaning similarity & Score from a model based on BERT and trained to predict human ratings of how well a text matches a reference \\
    \bottomrule
  \end{tabular}
\end{table}
 
\subsection{Model Selection}
 
Human and LLM Mental Health Conversations pairs questions from users with reference answers.
We used 3{,}307 examples for training and 200 for evaluation, split at random with a fixed seed.
We fine-tuned Gemma 3 with 4B and 27B parameters, Qwen 3 with 30B parameters and GPT-oss with 20B parameters.
We also evaluated GPT-oss with 120B parameters and DeepSeek V3 without fine-tuning.
Fine-tuning raised the overlap scores of most models but lowered BLEURT for all of them (Table~\ref{tab:selection}).
Gemma 3 27B changed little on any metric.
When we read the outputs, the other fine-tuned models produced replies that were unresponsive, incoherent or at times inappropriate.
 
\subsection{Training Corpus}
 
Requests for data from two earlier caregiver studies were unsuccessful, which led us to generate the corpus.
Its gender balance corrects the skew toward women that Dang's thesis reports as a limitation \cite{dang2024young}.
We removed exact and near duplicates with MinHash, a method that estimates how much two texts overlap.
Pairs with an estimated Jaccard similarity of 0.70 or more counted as near duplicates.
An audit confirmed that the assistant always played the supportive companion.
A rule-based pass then removed punctuation habits typical of LLM output, such as em dashes.
A second audit found no near duplicates between the training and held-out sets.
The rebuilt examples with retrieved context left the other 8{,}450 training examples and the held-out split unchanged.
 
\subsection{Retrieval Store}
 
The first store held 1{,}347 vectors, made of 1{,}293 document chunks and the 54 interview passages.
For the interview passages, the caregiver's words form the retrievable text.
One career guide supplied 1{,}019 of the 1{,}293 document chunks, about 76\%.
Many retrievals therefore returned general career advice whatever the user had asked.
The second store capped this guide at 300 chunks, for 771 vectors in total.
It added five sources on respite care, self-care, dementia behaviour, the emotional side of caregiving and long-distance caregiving.
All retrieval conditions reported in this paper use labeled passages.
 
\subsection{Self-Retrieval in Evaluation}
 
Gated retrieval at first appeared to raise the single-response score of Carebot by 0.16 points.
The 54 real evaluation inputs are also passages in the store, so retrieval often returned the user's own message.
After we excluded these self-matches, the apparent gain shrank to 0.04 points.
Production users are not passages in the store, so this artifact affects evaluation and not the deployed system.
 
\subsection{Tables}
Table~\ref{tab:participants} reports the age, gender, occupation and use of AI chatbots of the 90 participants in the analysis set, by condition.
Table~\ref{tab:selection} reports the model selection experiment.
Table~\ref{tab:refmetrics} reports the reference-based metrics for the four conditions on the 500 synthetic held-out examples.

We report these scores for completeness, since reply length confounds them.
 
\begin{table}[h]
  \caption{Model selection on Human and LLM Mental Health Conversations. Parameters give model size. Higher is better for every metric.}
  \label{tab:selection}
  \small
  \begin{tabular}{lcccccc}
    \toprule
    Model & Parameters & BERTScore & BLEU & ROUGE-L & METEOR & BLEURT \\
    \midrule
    Gemma 3, base & 4B & 0.5014 & 0.0020 & 0.0970 & 0.2384 & 0.4708 \\
    Gemma 3, fine-tuned & 4B & 0.5297 & 0.0035 & 0.1491 & 0.2122 & 0.4493 \\
    Gemma 3, base & 27B & 0.5023 & 0.0029 & 0.0980 & 0.2439 & 0.4633 \\
    Gemma 3, fine-tuned & 27B & 0.5020 & 0.0024 & 0.0982 & 0.2427 & 0.4627 \\
    Qwen 3, base & 30B & 0.4984 & 0.0020 & 0.1010 & 0.2361 & 0.4666 \\
    Qwen 3, fine-tuned & 30B & 0.5352 & 0.0039 & 0.1535 & 0.1979 & 0.4644 \\
    GPT-oss, base & 20B & 0.4720 & 0.0014 & 0.0875 & 0.2219 & 0.4238 \\
    GPT-oss, fine-tuned & 20B & 0.4944 & 0.0005 & 0.1136 & 0.2031 & 0.3427 \\
    GPT-oss, base & 120B & 0.4780 & 0.0011 & 0.0761 & 0.2142 & 0.4159 \\
    DeepSeek V3, base & 671B & 0.5176 & 0.0050 & 0.1240 & 0.2480 & 0.4772 \\
    \bottomrule
  \end{tabular}
\end{table}

\end{document}